\documentclass[aps,prb,reprint,superscriptaddress]{revtex4-1}

\usepackage{graphicx}
\usepackage{SIunits}
\usepackage{hyperref}

\begin{document}

\title{Magnon based logic in a multi-terminal YIG/Pt nanostructure}

\author{Kathrin Ganzhorn}
\email{kathrin.ganzhorn@wmi.badw.de}
\affiliation{Walther-Mei\ss{}ner-Institut, Bayerische Akademie der Wissenschaften, 85748 Garching, Germany}
\affiliation{Physik-Department, Technische Universit\"at M\"unchen, 85748 Garching, Germany}
\author{Stefan Klingler}
\affiliation{Walther-Mei\ss{}ner-Institut, Bayerische Akademie der Wissenschaften, 85748 Garching, Germany}
\affiliation{Physik-Department, Technische Universit\"at M\"unchen, 85748 Garching, Germany}
\author{Tobias Wimmer}
\affiliation{Walther-Mei\ss{}ner-Institut, Bayerische Akademie der Wissenschaften, 85748 Garching, Germany}
\affiliation{Physik-Department, Technische Universit\"at M\"unchen, 85748 Garching, Germany}
\author{Stephan Gepr{\"a}gs}
\affiliation{Walther-Mei\ss{}ner-Institut, Bayerische Akademie der Wissenschaften, 85748 Garching, Germany}
\author{Rudolf Gross}
\affiliation{Walther-Mei\ss{}ner-Institut, Bayerische Akademie der Wissenschaften, 85748 Garching, Germany}
\affiliation{Physik-Department, Technische Universit\"at M\"unchen, 85748 Garching, Germany}
\affiliation{Nanosystem Initiative Munich, 80799 M\"unchen, Germany}
\author{Hans Huebl}
\email{huebl@wmi.badw.de}
\affiliation{Walther-Mei\ss{}ner-Institut, Bayerische Akademie der Wissenschaften, 85748 Garching, Germany}
\affiliation{Physik-Department, Technische Universit\"at M\"unchen, 85748 Garching, Germany}
\affiliation{Nanosystem Initiative Munich, 80799 M\"unchen, Germany}
\author{Sebastian T. B. Goennenwein}
\email{goennenwein@wmi.badw.de}
\affiliation{Walther-Mei\ss{}ner-Institut, Bayerische Akademie der Wissenschaften, 85748 Garching, Germany}
\affiliation{Physik-Department, Technische Universit\"at M\"unchen, 85748 Garching, Germany}
\affiliation{Nanosystem Initiative Munich, 80799 M\"unchen, Germany}

\begin{abstract}

Boolean logic is the foundation of modern digital information processing. Recently, there has been a growing interest in phenomena based on pure spin currents, which allow to move from charge to spin based logic gates. We study a proof-of-principle logic device based on the ferrimagnetic insulator Yttrium Iron Garnet (YIG), with Pt strips acting as injectors and detectors for non-equilibrium magnons. We experimentally observe incoherent superposition of magnons generated by different injectors. This allows to implement a fully functional majority gate, enabling multiple logic operations (AND and OR) in one and the same device. Clocking frequencies of the order of several GHz and straightforward down-scaling make our device promising for applications.

\end{abstract}

\keywords{}

\maketitle

In the field of spintronics, the generation and detection of spin (polarized) currents have been studied extensively \cite{SaitohISHE, TserkovnyakSP, MosendzSP, CzeschkaSP, UchidaSSE, WeilerSSE, Althammer, Nakayama}, with the aim to use the spin degree of freedom to improve electronic devices.
In particular, magnonic devices based not on the flow of electrical charges but on the flow of quantized excitations of the spin system (magnons) have been considered for the transmission and processing of information. In this context, different propositions for the implementation of magnonic logic elements and their integration into the present electronic circuits have been made \cite{Kostylev2005, Schneider2008, Khitun, Chumak2014, Wagner2016}, using e.g.~the phase or the amplitude of magnon modes to encode a logical bit. 
One important step towards magnonic logic is the implementation of a majority gate \cite{Khitun}, with three inputs and one output. The output of the majority gate assumes the same logical state ("0 or "1") as the majority of the input signals. Additionally, one of the input channels can be used as a control channel to switch between "OR" and "AND" operations (see Tab.~\ref{majority_gate}), allowing for multiple types of logic operations in a single structure.  
Recently, a design for an all magnonic majority gate based on single magnon mode operations has been proposed \cite{ Klingler2014, Klingler2015}, using the phase of coherent spin waves (magnons) to encode the logical "1" and "0". However, in order to realize this magnonic majority gate, phase sensitive generation and detection of the spin waves is required. Furthermore, the selected magnon modes and therefore the functionality of the logic gate depend crucially on the waveguide geometry, making down-scaling challenging.
                          
\begin{figure}[!h]
\includegraphics[width=\columnwidth]{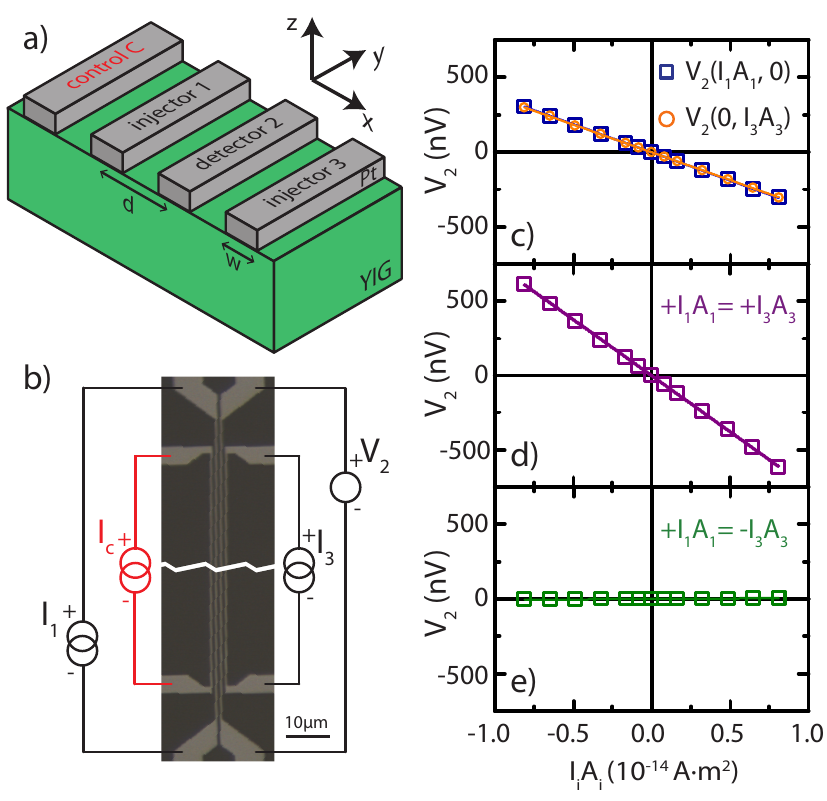}%
\caption{(a) Schematic of the YIG/Pt nanostructures consisting of four Pt strips with width $w$ and center-to-center separation $d$ . We label the strips as C (control), 1 (injector), 2 (detector) and 3 (injector) from left to right. (b) Optical micrograph of the YIG/Pt device: the bright strips are the Pt strips, the dark parts the YIG.  Current sources are attached to the injector strips 1 and 3 as well as to the control strip. The ensuing non-local voltage drop $V_2$ is recorded at strip 2. (c) The non-local voltage $V_2(I_1 A_1,0)$ while current-biasing only along strip 1, and $ V_2(0,I_3 A_3)$ while driving strip 3 are measured as a function of the applied current and are represented by blue open squares and orange open circles, respectively. The blue and orange lines represent linear fits to data. Note that we quote $I \cdot A$ as the relevant bias, since the spin Hall current across the entire interface area $A$ contributes to the MMR. (d) $V_2(I_1 A_1,I_3 A_3)$ measured while current-biasing strips 1 and 3 with the same current magnitude and polarity is represented by the purple open squares. The purple line is a linear fit. (e) The green open squares represent $V_2$ measured when biasing strips 1 and 3 with currents of opposite polarity. In this configuration, the non-local voltage goes to 0 within experimental error.}%
\label{data}%
\end{figure}

In this letter, we present a proof-of-principle device implementing a multi-terminal magnonic majority gate based on the incoherent superposition of magnons, requiring neither microwave excitation nor phase sensitive detection of the spin waves. For the implementation, we exploit the recently discovered magnon mediated magnetoresistance (MMR) \cite{Goennenwein2015, Cornelissen2015}, which was first measured in two parallel Pt strips separated by a distance $d$ of a few 100\,\nano\meter, deposited onto an Yttrium Iron Garnet (Y$_3$Fe$_5$O$_{12}$, YIG) thin film. By driving a charge current through the first strip, i.e. the injector, an electron spin accumulation  proportional to the driving current is generated in the Pt at the interface via the spin Hall effect. This spin accumulation is then converted into a magnon accumulation in the YIG beneath the injector \cite{ZhangPRB, ZhangPRL}, which diffuses across the ferrimagnetic insulator. We assume that the generated non-equilibrium magnons have the same polarization direction as the initial spin accumulation in the injector strip. If the second Pt strip, i.e. the detector, is within the magnon diffusion length, the non-equilibrium magnons in YIG are converted back into an electron spin accumulation in the detector Pt strip. This spin accumulation induces a charge current via the inverse spin Hall effect (ISHE), which is detected as non-local voltage in open circuit conditions.
Here, we extend the basic two Pt strip structure to a device with three magnon injectors and one detector as shown schematically in Fig.~\ref{data} (a). We observe incoherent superposition of magnons generated by different injector strips, i.e. the phase of the excited magnons is not relevant. We show that the detected non-local ISHE voltage is sensitive to the number and polarization of all non-equilibrium magnons accumulating beneath the detector. Based on these findings we are able to experimentally implement a four-strip magnon majority gate, which is integrated into an electronic circuit.

The YIG/Pt four-strip device we studied was fabricated starting from a commercially available $2\,\micro\meter$ thick YIG film grown onto 111 oriented Gd$_3$Ga$_5$O$_{12}$ via liquid phase epitaxy. After Piranha cleaning and annealing of the YIG to improve the interface quality (see Ref.\citenum{Goennenwein2015} for details), a $10\,\nano\meter$ thick Pt film was deposited on top of the YIG using electron beam evaporation. For the magnon injection and detection, we patterned four Pt strips with a width of $w=500\,\nano\meter$ and a center-to-center separation of $d = 1\,\micro\meter$ using electron beam lithography followed by Ar ion etching. An optical micrograph of the final device is shown in Fig.~\ref{data} (b). The two center strips (strip 1 and 2 in Fig.~\ref{data} (b)) have a length of $l=162\,\micro\meter$ and the outer strips (strip C on the left and strip 3 on the right) are $148\,\micro\meter$ long. The wiring scheme for the measurements is also sketched in Fig.~\ref{data} (b). All experiments are performed at $T = 275\,\kelvin$  with an external magnetic field $\mu_0H = 1\,\tesla$ applied in the thin film plane perpendicular to the strips. 

In the first part of the experiments we demonstrate the incoherent superposition of magnons in the simple three-strip device with strips 1, 2, and 3 (black wiring in Fig.~\ref{data} (b)): for the generation of the magnon spin accumulation beneath strip 1 and 3 we drive the charge currents $I_1$ and $I_3$ through the respective strips. As the number of magnons generated beneath one Pt strip is proportional to the electron spin accumulation at the interface, which in turn scales with the charge current $I$ flowing in the strip times the area $A$ of the YIG/Pt interface \cite{ZhangPRB}, the different lengths of injector strips 1 and 3 need to be taken into account for quantitative analysis. In order to detect the spin accumulation beneath strip 2, we therefore measure the open-circuit voltage $V_2(I_1 A_1, I_3 A_3)$ using a nanovoltmeter. To increase the measurement precision, we use the current switching method described in Ref.~\citenum{Goennenwein2015}.

Firstly, we focus on measuring the voltage response $V_2(I_1 A_1, I_3 A_3)$ while applying a current bias only to strip 1 ($I_1 \neq 0, I_3=0$) or to strip 3 ($I_1 = 0, I_3 \neq 0$), in order to compare the injection efficiency of both strips. The results are shown in Fig.~\ref{data} (c). As expected, $V_2$ scales linearly with $I_1$ and $I_3$, respectively, since the non-local voltage is proportional to the number of non-equilibrium magnons accumulating beneath strip 2 \cite{ZhangPRB, ZhangPRL}. Since at zero bias current no magnons are injected, $V_2(0, 0)=0$. For positive driving currents along the $+y$-direction in Fig.\ref{data} (a), the spin accumulation in the Pt is polarized along the $+x$-direction and thus we assume the magnons generated beneath the injector to be polarized along the same direction. These magnons diffuse to the detector (strip 2) and via the inverse spin Hall effect induce a negative voltage $V_2$, which is consistent with previous measurements using the same configuration\cite{Goennenwein2015}. By inverting the driving current, the spin viz.~magnon polarizations are inverted and consequently a positive non-local voltage is recorded. Taking these properties together, our detection method is therefore sensitive to the number of non-equilibrium magnons reaching the detector as well as their magnetic polarization. Note also that the magnons diffuse isotropically, i.e. to the left as well as to the right in Fig.~\ref{data} (a). Therefore, it does not matter for the sign of the detected voltage whether the injector strip is on the right or left side of the detector strip.

Within the experimental uncertainty of $5\,\nano\volt$ of our setup (limited by the nanovoltmeter noise and/or thermal stability of the setup), $V_2(I_1 A_1, 0) = V_2(0, I_3 A_3)$ for $I_1 A_1=I_3 A_3$, showing that the magnon generation efficiency of both strips is identical. To quantify this in more detail, we fit $V_2(I_1 A_1, 0)=\alpha_1 \cdot I_1 A_1$ and $V_2(0, I_3 A_3)=\alpha_3 \cdot I_3 A_3$ to the data. We find $\alpha_{1}=-3.77 \times 10^{7} \volt/\ampere\square\meter$ and $\alpha_{3}=-3.72 \times 10^{7} \volt/\ampere\square\meter$, showing that $\alpha_{1,3}$ are identical within less that 2\%.

Next, we investigate the non-local voltage $V_2(I_1 A_1, I_3 A_3)$ while simultaneously biasing strips 1 and 3. For identical current polarity ($I_1 A_1 = I_3 A_3$) equal numbers of magnons with identical polarization are generated beneath both strips. The results are shown in Fig.~\ref{data} (d) as purple open squares. The purple line represents a linear fit to the data with a slope of $\alpha_{1+3}=-7.53 \times 10^{7} \volt/\ampere\square\meter$. Assuming an incoherent superposition of the magnons created beneath strip 1 and strip 3 we expect $\alpha_{1+3}=\alpha_1+\alpha_3$, which is in good agreement with the experimental data. 
This result is further corroborated by the $V_2$ values obtained for opposite current directions in strip 1 and strip 3, i.e.~$I_1 A_1 = - I_3 A_3$, shown in Fig.~\ref{data} (e). A linear fit to the data reveals a slope of $0.05 \times 10^{7} \volt/\ampere\square\meter$ and therefore $\alpha_{3-1}=\alpha_3-\alpha_1$. Additional measurements were conducted using different driving current amplitudes and polarities for strips 1 and 3 (not shown here). From these data, we consistently find $V_2(I_1 A_1, I_3 A_3) = \alpha_1 \cdot I_1 A_1 + \alpha_3 \cdot I_3 A_3$. 

Taken together, we observe incoherent superposition of non-equilibrium magnons, injected independently by the different injector Pt strips.

\begin{table}[h]
	\begin{tabular}{|ccc||c|}
		\hline
			Control C & Injector 1 & Injector 3 & Detector 2\\
			\hline
			1 & 1 & 1 & 1\\
			1 & 0 & 1 & 1\\
			1 & 1 & 0 & 1\\
			1 & 0 & 0 & 0\\
			\hline
			0 & 1 & 1 & 1\\
			0 & 0 & 1 & 0\\
			0 & 1 & 0 & 0\\
			0 & 0 & 0 & 0\\
			\hline
			\hline
			$I_c$&$I_1$&$I_3$&$V_2$\\
			\hline
			$+$ & $+$ & $+$ & $-$\\
			$+$ & $+$ & $+$ & $-$\\
			$+$ & $+$ & $-$ & $-$\\
			$+$ & $-$ & $-$ & $+$\\
			\hline
			$-$ & $+$ & $+$ & $-$\\
			$-$ & $-$ & $+$ & $+$\\
			$-$ & $+$ & $-$ & $+$\\
			$-$ & $-$ & $-$ & $+$\\
			\hline
	\end{tabular}
	\caption{Truth table of a majority gate \cite{Klingler2014}. The bottom half of the table shows the sign of the bias currents for the input channels as well as the sign of the resulting non-local voltage $V_2$ in a MMR based majority gate. Positive and negative bias currents in the injectors are defined as logical "1" and "0" respectively. Since a positive bias current in the injector yields a negative non-local voltage $V_2$ \cite{Goennenwein2015}, we define $V_2 < 0$ as "1" and $V_2 > 0$ as "0".  }%
	\label{majority_gate}%
	
\end{table}

\begin{figure}[!h]
\includegraphics[width=\columnwidth]{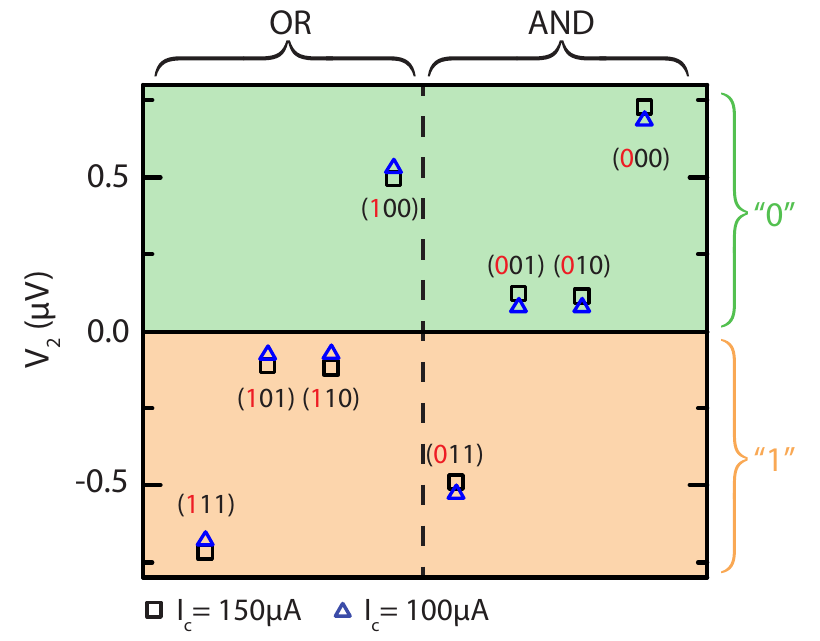}%
\caption{Test measurement of the four-strip majority gate. The detected non-local voltage $V_2$ is depicted for different input signals ($I_c~I_1~I_3$), where the bit of the control channel is marked in red. The magnitudes of the injector currents $I_1$ and $I_3$ are kept constant, and $V_2$ is measured for control current magnitudes $I_c=150$ and $100$ \,\micro\ampere~(black and blue symbols). The experimental data faithfully reproduce all the properties of a majority gate.}%
\label{majority_exp}%
\end{figure}

We now show that a magnon-based majority gate can be implemented in the four-strip nanostructure, with three input and one output channel as shown in Fig.~\ref{data} (a). A majority gate returns true if more than half the inputs are true, otherwise it returns false, as shown in Tab.~\ref{majority_gate}. The third input can be used as a so-called control channel (wire "C" in Fig.~\ref{data} (a)), which allows for switching between "AND" and "OR" operations. We define positive and negative currents in the injectors as the logical "1" and "0" respectively. Since for positive bias currents a negative non-local voltage is detected, for the output signal $V_2 < 0$ is defined as "1" and $V_2 > 0$ as "0". The bias currents for input 1 and 3 are chosen such that $\left|I_1 A_1\right|=\left|I_3 A_3\right|=0.81 \times 10^{-14} \ampere\square\meter$ and $V_2(I_1 A_1, I_3 A_3)$ is subsequently measured for fixed values of $I_c=\pm 150\,\micro\ampere$. The resulting voltage is shown as black open squares in Fig.~\ref{majority_exp} for different input combinations ($I_c~I_1~I_3$). For convenience, we do not quote the real charge current values, but rather the corresponding bit values (0 or 1), with the control bit marked in red. For example, if all three injectors are biased with a positive current, corresponding to the input (111), a negative voltage is detected at the output, corresponding to a logical "1". The experimental data successfully reproduces the truth table (Tab.~\ref{majority_gate}): by changing the control bit from "1" to "0", one can switch from an "OR" to an "AND" operation (left and right side of the graph respectively). The majority function is furthermore visible in the sign of the output signal (green and orange area in Fig.~\ref{majority_exp}), as the output mirrors the majority of the input signals.   

Since the control channel is further away from the detector, due to the exponential decay of the magnon accumulation with distance \cite{Goennenwein2015, Cornelissen2015}, the number of magnons from the control channel reaching the detector is smaller by about a factor 5.
However, the actual amplitude of the control signal $I_c$ is not crucial for the function of the majority gate and can be further reduced as shown in Fig.~\ref{majority_exp} for $100\,\micro\ampere$ (blue triangles): the majority gate function is not perturbed by changing the amplitude of the control bias current. 
We can therefore choose any amplitude for $I_c$ as long as the induced ISHE voltage is detectable. This allows for a high flexibility in the device operation. In this context, another advantage of this magnon based logic gate is the possibility to reprogram the device by simply interchanging the injector, detector and control channels. To guarantee the functionality of the majority gate, the input bias currents only need to be adapted to the geometry of the device, in particular to the distance between the injector and detector strips. 

Note that logic operations are already feasible in a three strip device with two injectors and one detector. In this case the threshold chosen for the detector determines whether the gate performs "AND" or "OR" operations. However, the four strip device allows for the same functions, without requiring a redefinition of the output threshold.

Apart from flexibility, an important aspect for the application of spintronic logic is the clocking frequency. The relevant time scale for a MMR based majority gate is dominated by two processes: the generation of a spin/magnon accumulation beneath the injector and the diffusion of the incoherent magnons. It has been shown experimentally that the spin Hall effect induced spin accumulation persists up to frequencies of at least a few GHz \cite{LotzeSMZ}, corresponding to spin accumulation build up times well below a nanosecond. Concerning the propagation of the spin waves, the average group velocity of magnons in YIG is of the order of 1\,\micro\meter/\nano\second \cite{Chumak2014}. In the device shown in Fig.~\ref{data}, the magnons therefore have a lower limit of about $2\,\nano\second$ for their travel time from the control to the detector strip, resulting in a maximum switching frequency $f=1/T$ of about $500\,\mega\hertz$. This can be further improved by changing the design of the four-strip device. For example reducing the width of the Pt strips to $50\nano\meter$ and their center-to-center separation to $100\nano\meter$ already leads to a factor 10 increase of the switching frequency, to several \,\giga\hertz. This value is comparable to clocking frequencies in current devices and to those expected in other spintronic based logic gates \cite{Stamps}. Down-scaling not only leads to faster switching, but also decreases the device footprint, and increases the energy effciency of the logic gate: for smaller distances between strips, the output signal increases exponentially \cite{Cornelissen2015}, such that the required bias currents are lower. Since the logical bit is encoded in the polarization of the generated magnons and not in the frequency, phase or amplitude of the spin waves, down-scaling does not perturb the functionality of the logic gate. Note that in the short distance limit, assuming ballistic magnon transport and neglecting interface losses, the non-local voltage is limited by the bias voltage times $\alpha_\mathrm{H}^2$, where $\alpha_\mathrm{H}$ is the spin Hall angle in the normal metal. This corresponds to the conversion efficiency from charge to spin and back to a charge current in the Pt layer. 
 
In summary, we have measured the non-local voltage (the magnon mediated magnetoresistance) in a YIG/Pt device with multiple magnon injectors and observe incoherent superposition of the non-equilibrium magnon populations. The measurements show that the detected non-local voltage is sensitive to the number of magnons reaching the detector and their polarization. Based on the incoherent superposition of spin waves, we implemented a fully functional four-strip majority gate. The logical bit in this device is encoded in the polarization of the magnons, which are injected using a dc charge current. The output can be read out as a dc non-local voltage, enabling a simple integration into an electronic circuit. Clocking frequencies of the order of several GHz and straightforward down-scaling make the device promising for applications. 

This work is financially supported by the Deutsche Forschungsgemeinschaft through the Priority Programm Spin Caloric Transport (GO 944/4).

\bibliography{bibliography}

\begin{thebibliography}{21}%
\makeatletter
\providecommand \@ifxundefined [1]{%
 \@ifx{#1\undefined}
}%
\providecommand \@ifnum [1]{%
 \ifnum #1\expandafter \@firstoftwo
 \else \expandafter \@secondoftwo
 \fi
}%
\providecommand \@ifx [1]{%
 \ifx #1\expandafter \@firstoftwo
 \else \expandafter \@secondoftwo
 \fi
}%
\providecommand \natexlab [1]{#1}%
\providecommand \enquote  [1]{``#1''}%
\providecommand \bibnamefont  [1]{#1}%
\providecommand \bibfnamefont [1]{#1}%
\providecommand \citenamefont [1]{#1}%
\providecommand \href@noop [0]{\@secondoftwo}%
\providecommand \href [0]{\begingroup \@sanitize@url \@href}%
\providecommand \@href[1]{\@@startlink{#1}\@@href}%
\providecommand \@@href[1]{\endgroup#1\@@endlink}%
\providecommand \@sanitize@url [0]{\catcode `\\12\catcode `\$12\catcode
  `\&12\catcode `\#12\catcode `\^12\catcode `\_12\catcode `\%12\relax}%
\providecommand \@@startlink[1]{}%
\providecommand \@@endlink[0]{}%
\providecommand \url  [0]{\begingroup\@sanitize@url \@url }%
\providecommand \@url [1]{\endgroup\@href {#1}{\urlprefix }}%
\providecommand \urlprefix  [0]{URL }%
\providecommand \Eprint [0]{\href }%
\providecommand \doibase [0]{http://dx.doi.org/}%
\providecommand \selectlanguage [0]{\@gobble}%
\providecommand \bibinfo  [0]{\@secondoftwo}%
\providecommand \bibfield  [0]{\@secondoftwo}%
\providecommand \translation [1]{[#1]}%
\providecommand \BibitemOpen [0]{}%
\providecommand \bibitemStop [0]{}%
\providecommand \bibitemNoStop [0]{.\EOS\space}%
\providecommand \EOS [0]{\spacefactor3000\relax}%
\providecommand \BibitemShut  [1]{\csname bibitem#1\endcsname}%
\let\auto@bib@innerbib\@empty
\bibitem [{\citenamefont {Saitoh}\ \emph {et~al.}(2006)\citenamefont {Saitoh},
  \citenamefont {Ueda}, \citenamefont {Miyajima},\ and\ \citenamefont
  {Tatara}}]{SaitohISHE}%
  \BibitemOpen
  \bibfield  {author} {\bibinfo {author} {\bibfnamefont {E.}~\bibnamefont
  {Saitoh}}, \bibinfo {author} {\bibfnamefont {M.}~\bibnamefont {Ueda}},
  \bibinfo {author} {\bibfnamefont {H.}~\bibnamefont {Miyajima}}, \ and\
  \bibinfo {author} {\bibfnamefont {G.}~\bibnamefont {Tatara}},\ }\href
  {\doibase http://dx.doi.org/10.1063/1.2199473} {\bibfield  {journal}
  {\bibinfo  {journal} {Appl. Phys. Lett.}\ }\textbf {\bibinfo {volume} {88}},\
  \bibinfo {eid} {182509} (\bibinfo {year} {2006}),\
  http://dx.doi.org/10.1063/1.2199473}\BibitemShut {NoStop}%
\bibitem [{\citenamefont {Tserkovnyak}\ \emph {et~al.}(2002)\citenamefont
  {Tserkovnyak}, \citenamefont {Brataas},\ and\ \citenamefont
  {Bauer}}]{TserkovnyakSP}%
  \BibitemOpen
  \bibfield  {author} {\bibinfo {author} {\bibfnamefont {Y.}~\bibnamefont
  {Tserkovnyak}}, \bibinfo {author} {\bibfnamefont {A.}~\bibnamefont
  {Brataas}}, \ and\ \bibinfo {author} {\bibfnamefont {G.~E.~W.}\ \bibnamefont
  {Bauer}},\ }\href {\doibase 10.1103/PhysRevLett.88.117601} {\bibfield
  {journal} {\bibinfo  {journal} {Phys. Rev. Lett.}\ }\textbf {\bibinfo
  {volume} {88}},\ \bibinfo {pages} {117601} (\bibinfo {year}
  {2002})}\BibitemShut {NoStop}%
\bibitem [{\citenamefont {Mosendz}\ \emph {et~al.}(2010)\citenamefont
  {Mosendz}, \citenamefont {Pearson}, \citenamefont {Fradin}, \citenamefont
  {Bauer}, \citenamefont {Bader},\ and\ \citenamefont {Hoffmann}}]{MosendzSP}%
  \BibitemOpen
  \bibfield  {author} {\bibinfo {author} {\bibfnamefont {O.}~\bibnamefont
  {Mosendz}}, \bibinfo {author} {\bibfnamefont {J.~E.}\ \bibnamefont
  {Pearson}}, \bibinfo {author} {\bibfnamefont {F.~Y.}\ \bibnamefont {Fradin}},
  \bibinfo {author} {\bibfnamefont {G.~E.~W.}\ \bibnamefont {Bauer}}, \bibinfo
  {author} {\bibfnamefont {S.~D.}\ \bibnamefont {Bader}}, \ and\ \bibinfo
  {author} {\bibfnamefont {A.}~\bibnamefont {Hoffmann}},\ }\href {\doibase
  10.1103/PhysRevLett.104.046601} {\bibfield  {journal} {\bibinfo  {journal}
  {Phys. Rev. Lett.}\ }\textbf {\bibinfo {volume} {104}},\ \bibinfo {pages}
  {046601} (\bibinfo {year} {2010})}\BibitemShut {NoStop}%
\bibitem [{\citenamefont {Czeschka}\ \emph {et~al.}(2011)\citenamefont
  {Czeschka}, \citenamefont {Dreher}, \citenamefont {Brandt}, \citenamefont
  {Weiler}, \citenamefont {Althammer}, \citenamefont {Imort}, \citenamefont
  {Reiss}, \citenamefont {Thomas}, \citenamefont {Schoch}, \citenamefont
  {Limmer}, \citenamefont {Huebl}, \citenamefont {Gross},\ and\ \citenamefont
  {Goennenwein}}]{CzeschkaSP}%
  \BibitemOpen
  \bibfield  {author} {\bibinfo {author} {\bibfnamefont {F.~D.}\ \bibnamefont
  {Czeschka}}, \bibinfo {author} {\bibfnamefont {L.}~\bibnamefont {Dreher}},
  \bibinfo {author} {\bibfnamefont {M.~S.}\ \bibnamefont {Brandt}}, \bibinfo
  {author} {\bibfnamefont {M.}~\bibnamefont {Weiler}}, \bibinfo {author}
  {\bibfnamefont {M.}~\bibnamefont {Althammer}}, \bibinfo {author}
  {\bibfnamefont {I.-M.}\ \bibnamefont {Imort}}, \bibinfo {author}
  {\bibfnamefont {G.}~\bibnamefont {Reiss}}, \bibinfo {author} {\bibfnamefont
  {A.}~\bibnamefont {Thomas}}, \bibinfo {author} {\bibfnamefont
  {W.}~\bibnamefont {Schoch}}, \bibinfo {author} {\bibfnamefont
  {W.}~\bibnamefont {Limmer}}, \bibinfo {author} {\bibfnamefont
  {H.}~\bibnamefont {Huebl}}, \bibinfo {author} {\bibfnamefont
  {R.}~\bibnamefont {Gross}}, \ and\ \bibinfo {author} {\bibfnamefont
  {S.~T.~B.}\ \bibnamefont {Goennenwein}},\ }\href {\doibase
  10.1103/PhysRevLett.107.046601} {\bibfield  {journal} {\bibinfo  {journal}
  {Phys. Rev. Lett.}\ }\textbf {\bibinfo {volume} {107}},\ \bibinfo {pages}
  {046601} (\bibinfo {year} {2011})}\BibitemShut {NoStop}%
\bibitem [{\citenamefont {Uchida}\ \emph {et~al.}(2010)\citenamefont {Uchida},
  \citenamefont {Adachi}, \citenamefont {Ota}, \citenamefont {Nakayama},
  \citenamefont {Maekawa},\ and\ \citenamefont {Saitoh}}]{UchidaSSE}%
  \BibitemOpen
  \bibfield  {author} {\bibinfo {author} {\bibfnamefont {K.}~\bibnamefont
  {Uchida}}, \bibinfo {author} {\bibfnamefont {H.}~\bibnamefont {Adachi}},
  \bibinfo {author} {\bibfnamefont {T.}~\bibnamefont {Ota}}, \bibinfo {author}
  {\bibfnamefont {H.}~\bibnamefont {Nakayama}}, \bibinfo {author}
  {\bibfnamefont {S.}~\bibnamefont {Maekawa}}, \ and\ \bibinfo {author}
  {\bibfnamefont {E.}~\bibnamefont {Saitoh}},\ }\href {\doibase
  10.1063/1.3507386} {\bibfield  {journal} {\bibinfo  {journal} {Appl. Phys.
  Lett.}\ }\textbf {\bibinfo {volume} {97}},\ \bibinfo {eid} {172505} (\bibinfo
  {year} {2010})}\BibitemShut {NoStop}%
\bibitem [{\citenamefont {Weiler}\ \emph {et~al.}(2012)\citenamefont {Weiler},
  \citenamefont {Althammer}, \citenamefont {Czeschka}, \citenamefont {Huebl},
  \citenamefont {Wagner}, \citenamefont {Opel}, \citenamefont {Imort},
  \citenamefont {Reiss}, \citenamefont {Thomas}, \citenamefont {Gross},\ and\
  \citenamefont {Goennenwein}}]{WeilerSSE}%
  \BibitemOpen
  \bibfield  {author} {\bibinfo {author} {\bibfnamefont {M.}~\bibnamefont
  {Weiler}}, \bibinfo {author} {\bibfnamefont {M.}~\bibnamefont {Althammer}},
  \bibinfo {author} {\bibfnamefont {F.~D.}\ \bibnamefont {Czeschka}}, \bibinfo
  {author} {\bibfnamefont {H.}~\bibnamefont {Huebl}}, \bibinfo {author}
  {\bibfnamefont {M.~S.}\ \bibnamefont {Wagner}}, \bibinfo {author}
  {\bibfnamefont {M.}~\bibnamefont {Opel}}, \bibinfo {author} {\bibfnamefont
  {I.-M.}\ \bibnamefont {Imort}}, \bibinfo {author} {\bibfnamefont
  {G.}~\bibnamefont {Reiss}}, \bibinfo {author} {\bibfnamefont
  {A.}~\bibnamefont {Thomas}}, \bibinfo {author} {\bibfnamefont
  {R.}~\bibnamefont {Gross}}, \ and\ \bibinfo {author} {\bibfnamefont
  {S.~T.~B.}\ \bibnamefont {Goennenwein}},\ }\href {\doibase
  10.1103/PhysRevLett.108.106602} {\bibfield  {journal} {\bibinfo  {journal}
  {Phys. Rev. Lett.}\ }\textbf {\bibinfo {volume} {108}},\ \bibinfo {pages}
  {106602} (\bibinfo {year} {2012})}\BibitemShut {NoStop}%
\bibitem [{\citenamefont {Althammer}\ \emph {et~al.}(2013)\citenamefont
  {Althammer}, \citenamefont {Meyer}, \citenamefont {Nakayama}, \citenamefont
  {Schreier}, \citenamefont {Altmannshofer}, \citenamefont {Weiler},
  \citenamefont {Huebl}, \citenamefont {Gepr\"ags}, \citenamefont {Opel},
  \citenamefont {Gross}, \citenamefont {Meier}, \citenamefont {Klewe},
  \citenamefont {Kuschel}, \citenamefont {Schmalhorst}, \citenamefont {Reiss},
  \citenamefont {Shen}, \citenamefont {Gupta}, \citenamefont {Chen},
  \citenamefont {Bauer}, \citenamefont {Saitoh},\ and\ \citenamefont
  {Goennenwein}}]{Althammer}%
  \BibitemOpen
  \bibfield  {author} {\bibinfo {author} {\bibfnamefont {M.}~\bibnamefont
  {Althammer}}, \bibinfo {author} {\bibfnamefont {S.}~\bibnamefont {Meyer}},
  \bibinfo {author} {\bibfnamefont {H.}~\bibnamefont {Nakayama}}, \bibinfo
  {author} {\bibfnamefont {M.}~\bibnamefont {Schreier}}, \bibinfo {author}
  {\bibfnamefont {S.}~\bibnamefont {Altmannshofer}}, \bibinfo {author}
  {\bibfnamefont {M.}~\bibnamefont {Weiler}}, \bibinfo {author} {\bibfnamefont
  {H.}~\bibnamefont {Huebl}}, \bibinfo {author} {\bibfnamefont
  {S.}~\bibnamefont {Gepr\"ags}}, \bibinfo {author} {\bibfnamefont
  {M.}~\bibnamefont {Opel}}, \bibinfo {author} {\bibfnamefont {R.}~\bibnamefont
  {Gross}}, \bibinfo {author} {\bibfnamefont {D.}~\bibnamefont {Meier}},
  \bibinfo {author} {\bibfnamefont {C.}~\bibnamefont {Klewe}}, \bibinfo
  {author} {\bibfnamefont {T.}~\bibnamefont {Kuschel}}, \bibinfo {author}
  {\bibfnamefont {J.-M.}\ \bibnamefont {Schmalhorst}}, \bibinfo {author}
  {\bibfnamefont {G.}~\bibnamefont {Reiss}}, \bibinfo {author} {\bibfnamefont
  {L.}~\bibnamefont {Shen}}, \bibinfo {author} {\bibfnamefont {A.}~\bibnamefont
  {Gupta}}, \bibinfo {author} {\bibfnamefont {Y.-T.}\ \bibnamefont {Chen}},
  \bibinfo {author} {\bibfnamefont {G.~E.~W.}\ \bibnamefont {Bauer}}, \bibinfo
  {author} {\bibfnamefont {E.}~\bibnamefont {Saitoh}}, \ and\ \bibinfo {author}
  {\bibfnamefont {S.~T.~B.}\ \bibnamefont {Goennenwein}},\ }\href {\doibase
  10.1103/PhysRevB.87.224401} {\bibfield  {journal} {\bibinfo  {journal} {Phys.
  Rev. B}\ }\textbf {\bibinfo {volume} {87}},\ \bibinfo {pages} {224401}
  (\bibinfo {year} {2013})}\BibitemShut {NoStop}%
\bibitem [{\citenamefont {Nakayama}\ \emph {et~al.}(2013)\citenamefont
  {Nakayama}, \citenamefont {Althammer}, \citenamefont {Chen}, \citenamefont
  {Uchida}, \citenamefont {Kajiwara}, \citenamefont {Kikuchi}, \citenamefont
  {Ohtani}, \citenamefont {Gepr\"ags}, \citenamefont {Opel}, \citenamefont
  {Takahashi}, \citenamefont {Gross}, \citenamefont {Bauer}, \citenamefont
  {Goennenwein},\ and\ \citenamefont {Saitoh}}]{Nakayama}%
  \BibitemOpen
  \bibfield  {author} {\bibinfo {author} {\bibfnamefont {H.}~\bibnamefont
  {Nakayama}}, \bibinfo {author} {\bibfnamefont {M.}~\bibnamefont {Althammer}},
  \bibinfo {author} {\bibfnamefont {Y.-T.}\ \bibnamefont {Chen}}, \bibinfo
  {author} {\bibfnamefont {K.}~\bibnamefont {Uchida}}, \bibinfo {author}
  {\bibfnamefont {Y.}~\bibnamefont {Kajiwara}}, \bibinfo {author}
  {\bibfnamefont {D.}~\bibnamefont {Kikuchi}}, \bibinfo {author} {\bibfnamefont
  {T.}~\bibnamefont {Ohtani}}, \bibinfo {author} {\bibfnamefont
  {S.}~\bibnamefont {Gepr\"ags}}, \bibinfo {author} {\bibfnamefont
  {M.}~\bibnamefont {Opel}}, \bibinfo {author} {\bibfnamefont {S.}~\bibnamefont
  {Takahashi}}, \bibinfo {author} {\bibfnamefont {R.}~\bibnamefont {Gross}},
  \bibinfo {author} {\bibfnamefont {G.~E.~W.}\ \bibnamefont {Bauer}}, \bibinfo
  {author} {\bibfnamefont {S.~T.~B.}\ \bibnamefont {Goennenwein}}, \ and\
  \bibinfo {author} {\bibfnamefont {E.}~\bibnamefont {Saitoh}},\ }\href
  {\doibase 10.1103/PhysRevLett.110.206601} {\bibfield  {journal} {\bibinfo
  {journal} {Phys. Rev. Lett.}\ }\textbf {\bibinfo {volume} {110}},\ \bibinfo
  {pages} {206601} (\bibinfo {year} {2013})}\BibitemShut {NoStop}%
\bibitem [{\citenamefont {Kostylev}\ \emph {et~al.}(2005)\citenamefont
  {Kostylev}, \citenamefont {Serga}, \citenamefont {Schneider}, \citenamefont
  {Leven},\ and\ \citenamefont {Hillebrands}}]{Kostylev2005}%
  \BibitemOpen
  \bibfield  {author} {\bibinfo {author} {\bibfnamefont {M.~P.}\ \bibnamefont
  {Kostylev}}, \bibinfo {author} {\bibfnamefont {A.~A.}\ \bibnamefont {Serga}},
  \bibinfo {author} {\bibfnamefont {T.}~\bibnamefont {Schneider}}, \bibinfo
  {author} {\bibfnamefont {B.}~\bibnamefont {Leven}}, \ and\ \bibinfo {author}
  {\bibfnamefont {B.}~\bibnamefont {Hillebrands}},\ }\href {\doibase
  10.1063/1.2089147} {\bibfield  {journal} {\bibinfo  {journal} {Appl. Phys.
  Lett.}\ }\textbf {\bibinfo {volume} {87}},\ \bibinfo {pages} {153501}
  (\bibinfo {year} {2005})}\BibitemShut {NoStop}%
\bibitem [{\citenamefont {Schneider}\ \emph {et~al.}(2008)\citenamefont
  {Schneider}, \citenamefont {Serga}, \citenamefont {Leven}, \citenamefont
  {Hillebrands}, \citenamefont {Stamps},\ and\ \citenamefont
  {Kostylev}}]{Schneider2008}%
  \BibitemOpen
  \bibfield  {author} {\bibinfo {author} {\bibfnamefont {T.}~\bibnamefont
  {Schneider}}, \bibinfo {author} {\bibfnamefont {A.~A.}\ \bibnamefont
  {Serga}}, \bibinfo {author} {\bibfnamefont {B.}~\bibnamefont {Leven}},
  \bibinfo {author} {\bibfnamefont {B.}~\bibnamefont {Hillebrands}}, \bibinfo
  {author} {\bibfnamefont {R.~L.}\ \bibnamefont {Stamps}}, \ and\ \bibinfo
  {author} {\bibfnamefont {M.~P.}\ \bibnamefont {Kostylev}},\ }\href {\doibase
  10.1063/1.2834714} {\bibfield  {journal} {\bibinfo  {journal} {Appl. Phys.
  Lett.}\ }\textbf {\bibinfo {volume} {92}},\ \bibinfo {pages} {22505}
  (\bibinfo {year} {2008})}\BibitemShut {NoStop}%
\bibitem [{\citenamefont {Khitun}\ \emph {et~al.}(2008)\citenamefont {Khitun},
  \citenamefont {Bao},\ and\ \citenamefont {Wang}}]{Khitun}%
  \BibitemOpen
  \bibfield  {author} {\bibinfo {author} {\bibfnamefont {A.}~\bibnamefont
  {Khitun}}, \bibinfo {author} {\bibfnamefont {M.}~\bibnamefont {Bao}}, \ and\
  \bibinfo {author} {\bibfnamefont {K.~L.}\ \bibnamefont {Wang}},\ }\href
  {\doibase 10.1109/TMAG.2008.2000812} {\bibfield  {journal} {\bibinfo
  {journal} {IEEE Trans. Magn.}\ }\textbf {\bibinfo {volume} {44}},\ \bibinfo
  {pages} {2141} (\bibinfo {year} {2008})}\BibitemShut {NoStop}%
\bibitem [{\citenamefont {Chumak}\ \emph {et~al.}(2014)\citenamefont {Chumak},
  \citenamefont {Serga},\ and\ \citenamefont {Hillebrands}}]{Chumak2014}%
  \BibitemOpen
  \bibfield  {author} {\bibinfo {author} {\bibfnamefont {A.~V.}\ \bibnamefont
  {Chumak}}, \bibinfo {author} {\bibfnamefont {A.~A.}\ \bibnamefont {Serga}}, \
  and\ \bibinfo {author} {\bibfnamefont {B.}~\bibnamefont {Hillebrands}},\
  }\href {\doibase 10.1038/ncomms5700} {\bibfield  {journal} {\bibinfo
  {journal} {Nature Communications}\ }\textbf {\bibinfo {volume} {5}},\
  \bibinfo {pages} {4700} (\bibinfo {year} {2014})}\BibitemShut {NoStop}%
\bibitem [{\citenamefont {Wagner}\ \emph {et~al.}(2016)\citenamefont {Wagner},
  \citenamefont {K\'{a}kay}, \citenamefont {Schultheiss}, \citenamefont
  {Henschke}, \citenamefont {Sebastian},\ and\ \citenamefont
  {Schultheiss}}]{Wagner2016}%
  \BibitemOpen
  \bibfield  {author} {\bibinfo {author} {\bibfnamefont {K.}~\bibnamefont
  {Wagner}}, \bibinfo {author} {\bibfnamefont {K.}~\bibnamefont {K\'{a}kay}},
  \bibinfo {author} {\bibfnamefont {K.}~\bibnamefont {Schultheiss}}, \bibinfo
  {author} {\bibfnamefont {A.}~\bibnamefont {Henschke}}, \bibinfo {author}
  {\bibfnamefont {T.}~\bibnamefont {Sebastian}}, \ and\ \bibinfo {author}
  {\bibfnamefont {H.}~\bibnamefont {Schultheiss}},\ }\href
  {http://dx.doi.org/10.1038/nnano.2015.339} {\bibfield  {journal} {\bibinfo
  {journal} {Nature Nanotechnology}\ }\textbf {\bibinfo {volume} {advance
  online publication}},\  (\bibinfo {year} {2016})}\BibitemShut {NoStop}%
\bibitem [{\citenamefont {Klingler}\ \emph {et~al.}(2014)\citenamefont
  {Klingler}, \citenamefont {Pirro}, \citenamefont {Br{\"{a}}cher},
  \citenamefont {Leven}, \citenamefont {Hillebrands},\ and\ \citenamefont
  {Chumak}}]{Klingler2014}%
  \BibitemOpen
  \bibfield  {author} {\bibinfo {author} {\bibfnamefont {S.}~\bibnamefont
  {Klingler}}, \bibinfo {author} {\bibfnamefont {P.}~\bibnamefont {Pirro}},
  \bibinfo {author} {\bibfnamefont {T.}~\bibnamefont {Br{\"{a}}cher}}, \bibinfo
  {author} {\bibfnamefont {B.}~\bibnamefont {Leven}}, \bibinfo {author}
  {\bibfnamefont {B.}~\bibnamefont {Hillebrands}}, \ and\ \bibinfo {author}
  {\bibfnamefont {A.~V.}\ \bibnamefont {Chumak}},\ }\href {\doibase
  10.1063/1.4898042} {\bibfield  {journal} {\bibinfo  {journal} {Appl. Phys.
  Lett.}\ }\textbf {\bibinfo {volume} {105}},\ \bibinfo {pages} {152410}
  (\bibinfo {year} {2014})}\BibitemShut {NoStop}%
\bibitem [{\citenamefont {Klingler}\ \emph {et~al.}(2015)\citenamefont
  {Klingler}, \citenamefont {Pirro}, \citenamefont {Br{\"{a}}cher},
  \citenamefont {Leven}, \citenamefont {Hillebrands},\ and\ \citenamefont
  {Chumak}}]{Klingler2015}%
  \BibitemOpen
  \bibfield  {author} {\bibinfo {author} {\bibfnamefont {S.}~\bibnamefont
  {Klingler}}, \bibinfo {author} {\bibfnamefont {P.}~\bibnamefont {Pirro}},
  \bibinfo {author} {\bibfnamefont {T.}~\bibnamefont {Br{\"{a}}cher}}, \bibinfo
  {author} {\bibfnamefont {B.}~\bibnamefont {Leven}}, \bibinfo {author}
  {\bibfnamefont {B.}~\bibnamefont {Hillebrands}}, \ and\ \bibinfo {author}
  {\bibfnamefont {A.~V.}\ \bibnamefont {Chumak}},\ }\href {\doibase
  10.1063/1.4921850} {\bibfield  {journal} {\bibinfo  {journal} {Appl. Phys.
  Lett.}\ }\textbf {\bibinfo {volume} {106}},\ \bibinfo {pages} {212406}
  (\bibinfo {year} {2015})}\BibitemShut {NoStop}%
\bibitem [{\citenamefont {Goennenwein}\ \emph {et~al.}(2015)\citenamefont
  {Goennenwein}, \citenamefont {Schlitz}, \citenamefont {Pernpeintner},
  \citenamefont {Ganzhorn}, \citenamefont {Althammer}, \citenamefont {Gross},\
  and\ \citenamefont {Huebl}}]{Goennenwein2015}%
  \BibitemOpen
  \bibfield  {author} {\bibinfo {author} {\bibfnamefont {S.~T.~B.}\
  \bibnamefont {Goennenwein}}, \bibinfo {author} {\bibfnamefont
  {R.}~\bibnamefont {Schlitz}}, \bibinfo {author} {\bibfnamefont
  {M.}~\bibnamefont {Pernpeintner}}, \bibinfo {author} {\bibfnamefont
  {K.}~\bibnamefont {Ganzhorn}}, \bibinfo {author} {\bibfnamefont
  {M.}~\bibnamefont {Althammer}}, \bibinfo {author} {\bibfnamefont
  {R.}~\bibnamefont {Gross}}, \ and\ \bibinfo {author} {\bibfnamefont
  {H.}~\bibnamefont {Huebl}},\ }\href {\doibase 10.1063/1.4935074} {\bibfield
  {journal} {\bibinfo  {journal} {Appl. Phys. Lett.}\ }\textbf {\bibinfo
  {volume} {107}},\ \bibinfo {pages} {172405} (\bibinfo {year}
  {2015})}\BibitemShut {NoStop}%
\bibitem [{\citenamefont {Cornelissen}\ \emph {et~al.}(2015)\citenamefont
  {Cornelissen}, \citenamefont {Liu}, \citenamefont {Duine}, \citenamefont
  {Youssef},\ and\ \citenamefont {van Wees}}]{Cornelissen2015}%
  \BibitemOpen
  \bibfield  {author} {\bibinfo {author} {\bibfnamefont {L.~J.}\ \bibnamefont
  {Cornelissen}}, \bibinfo {author} {\bibfnamefont {J.}~\bibnamefont {Liu}},
  \bibinfo {author} {\bibfnamefont {R.~A.}\ \bibnamefont {Duine}}, \bibinfo
  {author} {\bibfnamefont {J.~B.}\ \bibnamefont {Youssef}}, \ and\ \bibinfo
  {author} {\bibfnamefont {B.~J.}\ \bibnamefont {van Wees}},\ }\href@noop {}
  {\bibfield  {journal} {\bibinfo  {journal} {Nature Physics}\ }\textbf
  {\bibinfo {volume} {11}},\ \bibinfo {pages} {1022} (\bibinfo {year}
  {2015})}\BibitemShut {NoStop}%
\bibitem [{\citenamefont {Zhang}\ and\ \citenamefont
  {Zhang}(2012{\natexlab{a}})}]{ZhangPRB}%
  \BibitemOpen
  \bibfield  {author} {\bibinfo {author} {\bibfnamefont {S.~S.-L.}\
  \bibnamefont {Zhang}}\ and\ \bibinfo {author} {\bibfnamefont
  {S.}~\bibnamefont {Zhang}},\ }\href {\doibase 10.1103/PhysRevB.86.214424}
  {\bibfield  {journal} {\bibinfo  {journal} {Phys. Rev. B}\ }\textbf {\bibinfo
  {volume} {86}},\ \bibinfo {pages} {214424} (\bibinfo {year}
  {2012}{\natexlab{a}})}\BibitemShut {NoStop}%
\bibitem [{\citenamefont {Zhang}\ and\ \citenamefont
  {Zhang}(2012{\natexlab{b}})}]{ZhangPRL}%
  \BibitemOpen
  \bibfield  {author} {\bibinfo {author} {\bibfnamefont {S.~S.-L.}\
  \bibnamefont {Zhang}}\ and\ \bibinfo {author} {\bibfnamefont
  {S.}~\bibnamefont {Zhang}},\ }\href {\doibase 10.1103/PhysRevLett.109.096603}
  {\bibfield  {journal} {\bibinfo  {journal} {Phys. Rev. Lett.}\ }\textbf
  {\bibinfo {volume} {109}},\ \bibinfo {pages} {096603} (\bibinfo {year}
  {2012}{\natexlab{b}})}\BibitemShut {NoStop}%
\bibitem [{\citenamefont {Lotze}\ \emph {et~al.}(2014)\citenamefont {Lotze},
  \citenamefont {Huebl}, \citenamefont {Gross},\ and\ \citenamefont
  {Goennenwein}}]{LotzeSMZ}%
  \BibitemOpen
  \bibfield  {author} {\bibinfo {author} {\bibfnamefont {J.}~\bibnamefont
  {Lotze}}, \bibinfo {author} {\bibfnamefont {H.}~\bibnamefont {Huebl}},
  \bibinfo {author} {\bibfnamefont {R.}~\bibnamefont {Gross}}, \ and\ \bibinfo
  {author} {\bibfnamefont {S.~T.~B.}\ \bibnamefont {Goennenwein}},\ }\href
  {\doibase 10.1103/PhysRevB.90.174419} {\bibfield  {journal} {\bibinfo
  {journal} {Phys. Rev. B}\ }\textbf {\bibinfo {volume} {90}},\ \bibinfo
  {pages} {174419} (\bibinfo {year} {2014})}\BibitemShut {NoStop}%
\bibitem [{\citenamefont {Stamps}\ \emph {et~al.}(2014)\citenamefont {Stamps},
  \citenamefont {Breitkreutz}, \citenamefont {Akerman}, \citenamefont {Chumak},
  \citenamefont {Otani}, \citenamefont {Bauer}, \citenamefont {Thiele},
  \citenamefont {Bowen}, \citenamefont {Majetich}, \citenamefont {Kl\"aui},
  \citenamefont {Prejbeanu}, \citenamefont {Dieny}, \citenamefont {Dempsey},\
  and\ \citenamefont {Hillebrands}}]{Stamps}%
  \BibitemOpen
  \bibfield  {author} {\bibinfo {author} {\bibfnamefont {R.~L.}\ \bibnamefont
  {Stamps}}, \bibinfo {author} {\bibfnamefont {S.}~\bibnamefont {Breitkreutz}},
  \bibinfo {author} {\bibfnamefont {J.}~\bibnamefont {Akerman}}, \bibinfo
  {author} {\bibfnamefont {A.~V.}\ \bibnamefont {Chumak}}, \bibinfo {author}
  {\bibfnamefont {Y.}~\bibnamefont {Otani}}, \bibinfo {author} {\bibfnamefont
  {G.~E.~W.}\ \bibnamefont {Bauer}}, \bibinfo {author} {\bibfnamefont {J.-U.}\
  \bibnamefont {Thiele}}, \bibinfo {author} {\bibfnamefont {M.}~\bibnamefont
  {Bowen}}, \bibinfo {author} {\bibfnamefont {S.~A.}\ \bibnamefont {Majetich}},
  \bibinfo {author} {\bibfnamefont {M.}~\bibnamefont {Kl\"aui}}, \bibinfo
  {author} {\bibfnamefont {I.~L.}\ \bibnamefont {Prejbeanu}}, \bibinfo {author}
  {\bibfnamefont {B.}~\bibnamefont {Dieny}}, \bibinfo {author} {\bibfnamefont
  {N.~M.}\ \bibnamefont {Dempsey}}, \ and\ \bibinfo {author} {\bibfnamefont
  {B.}~\bibnamefont {Hillebrands}},\ }\href
  {http://stacks.iop.org/0022-3727/47/i=33/a=333001} {\bibfield  {journal}
  {\bibinfo  {journal} {J. Phys. D: Appl. Phys.}\ }\textbf {\bibinfo {volume}
  {47}},\ \bibinfo {pages} {333001} (\bibinfo {year} {2014})}\BibitemShut
  {NoStop}%
\end{thebibliography}%

\end{document}